%
%
%
\font\ninerm=cmr9
\font\eightrm=cmr8
\font\sixrm=cmr6
\font\ninei=cmmi9
\font\eighti=cmmi8
\font\sixi=cmmi6
\skewchar\ninei='177 \skewchar\eighti='177 \skewchar\sixi='177
\font\ninesy=cmsy9
\font\eightsy=cmsy8
\font\sixsy=cmsy6
\skewchar\ninesy='60 \skewchar\eightsy='60 \skewchar\sixsy='60

\font\ninebf=cmbx9
\font\eightbf=cmbx8
\font\sixbf=cmbx6
\font\ninett=cmtt9
\font\eighttt=cmtt8
\hyphenchar\tentt=-1 
\hyphenchar\ninett=-1
\hyphenchar\eighttt=-1
\font\ninesl=cmsl9
\font\eightsl=cmsl8
\font\nineit=cmti9
\font\eightit=cmti8
\newskip\ttglue
\def\tenpoint{\def\rm{\fam0\tenrm}%
  \textfont0=\tenrm \scriptfont0=\sevenrm \scriptscriptfont0=\fiverm
  \textfont1=\teni \scriptfont1=\seveni \scriptscriptfont1=\fivei
  \textfont2=\tensy \scriptfont2=\sevensy \scriptscriptfont2=\fivesy
  \textfont3=\tenex \scriptfont3=\tenex \scriptscriptfont3=\tenex
  \def\it{\fam\itfam\tenit}%
  \textfont\itfam=\tenit
  \def\sl{\fam\slfam\tensl}%
  \textfont\slfam=\tensl
  \def\bf{\fam\bffam\tenbf}%
  \textfont\bffam=\tenbf \scriptfont\bffam=\sevenbf
   \scriptscriptfont\bffam=\fivebf
  \def\tt{\fam\ttfam\tentt}%
  \textfont\ttfam=\tentt
  \tt \ttglue=.5em plus.25em minus.15em
  \normalbaselineskip=12pt
  \let\sc=\eightrm
  \let\big=\tenbig
  \setbox\strutbox=\hbox{\vrule height8.5pt depth3.5pt width0pt}%
  \normalbaselines\rm}
\def\ninepoint{\def\rm{\fam0\ninerm}%
  \textfont0=\ninerm \scriptfont0=\sixrm \scriptscriptfont0=\fiverm
  \textfont1=\ninei \scriptfont1=\sixi \scriptscriptfont1=\fivei
  \textfont2=\ninesy \scriptfont2=\sixsy \scriptscriptfont2=\fivesy
  \textfont3=\tenex \scriptfont3=\tenex \scriptscriptfont3=\tenex
  \def\it{\fam\itfam\nineit}%
  \textfont\itfam=\nineit
  \def\sl{\fam\slfam\ninesl}%
  \textfont\slfam=\ninesl
  \def\bf{\fam\bffam\ninebf}%
  \textfont\bffam=\ninebf \scriptfont\bffam=\sixbf
   \scriptscriptfont\bffam=\fivebf
  \def\tt{\fam\ttfam\ninett}%
  \textfont\ttfam=\ninett
  \tt \ttglue=.5em plus.25em minus.15em
  \normalbaselineskip=10pt 
  \let\sc=\sevenrm
  \let\big=\ninebig
  \setbox\strutbox=\hbox{\vrule height8pt depth3pt width0pt}%
  \normalbaselines\rm}
\def\eightpoint{\def\rm{\fam0\eightrm}%
  \textfont0=\eightrm \scriptfont0=\sixrm \scriptscriptfont0=\fiverm
  \textfont1=\eighti \scriptfont1=\sixi \scriptscriptfont1=\fivei
  \textfont2=\eightsy \scriptfont2=\sixsy \scriptscriptfont2=\fivesy
  \textfont3=\tenex \scriptfont3=\tenex \scriptscriptfont3=\tenex
  \def\it{\fam\itfam\eightit}%
  \textfont\itfam=\eightit
  \def\sl{\fam\slfam\eightsl}%
  \textfont\slfam=\eightsl
  \def\bf{\fam\bffam\eightbf}%
  \textfont\bffam=\eightbf \scriptfont\bffam=\sixbf
   \scriptscriptfont\bffam=\fivebf
  \def\tt{\fam\ttfam\eighttt}%
  \textfont\ttfam=\eighttt
  \tt \ttglue=.5em plus.25em minus.15em
  \normalbaselineskip=9pt
  \let\sc=\sixrm
  \let\big=\eightbig
  \setbox\strutbox=\hbox{\vrule height7pt depth2pt width0pt}%
  \normalbaselines\rm}
%
\def\headtype{\ninepoint}                 
\def\abstracttype{\ninepoint}             
\def\captiontype{\ninepoint}              
\def\footnotetype{\ninepoint}             
\def\refit{\it}                           
\font\chaptitle=cmr10 at 11pt             
\rm                                       

%
%
\parindent=0.25in                         
\parskip=0pt                              
\baselineskip=12pt                        
\hsize=4.25truein                         
\vsize=7.445truein                        
\hoffset=1in                              
\voffset=-0.5in                           

\newskip\sectionskipamount                
\newskip\aftermainskipamount              
\newskip\subsecskipamount                 
\newskip\firstpageskipamount              
\newskip\capskipamount                    
\newskip\ackskipamount                    
\sectionskipamount=0.2in plus 0.09in
\aftermainskipamount=6pt plus 6pt         
\subsecskipamount=0.1in plus 0.04in
\firstpageskipamount=3pc
\capskipamount=0.1in
\ackskipamount=0.15in
\def\sectionskip{\vskip\sectionskipamount}
\def\aftermainskip{\vskip\aftermainskipamount}
\def\subsecskip{\vskip\subsecskipamount} 
\def\firstpageskip{\vskip\firstpageskipamount}
\def\capskip{\hskip\capskipamount}

%
%
\nopagenumbers                            
\newcount\firstpageno                     
\firstpageno=\pageno                      
\newcount\chapno                          

\def\rightheadline{\headtype\phantom{\folio}\hfil\runningtitletext\hfil\folio}
\def\leftheadline{\headtype\folio\hfil\runningauthortext\hfil\phantom{\folio}}
\headline={\ifnum\pageno=\firstpageno\hfil
           \else
              \ifdim\ht\topins=\vsize           
                 \ifdim\dp\topins=1sp \hfil     
                 \else
                     \ifodd\pageno\rightheadline\else\leftheadline\fi
                 \fi
              \else
                 \ifodd\pageno\rightheadline\else\leftheadline\fi
              \fi
           \fi}

\def\bottomnumber{\hss\tenrm[\folio]\hss}
\footline={\ifnum\pageno=\firstpageno\bottomnumber\else\hfil\fi}

%
%
%
%
\outer\def\mainsection#1
    {\vskip 0pt plus\smallskipamount\sectionskip
     \message{#1}\vbox{\noindent{\bf#1}}\nobreak\aftermainskip\noindent}
 
\outer\def\subsection#1
    {\vskip 0pt plus\smallskipamount\subsecskip
     \message{#1}\vbox{\noindent{\bf#1}}\nobreak\smallskip\nobreak\noindent}
 
\def\backup{\nobreak\vskip-\baselineskip\nobreak\vskip-\subsecskipamount\nobreak
}

\def\title#1{{\chaptitle\leftline{#1}}}
\def\name#1{\leftline{#1}}
\def\affiliation#1{\leftline{\it #1}}
\def\abstract#1{{\abstracttype \noindent #1 \smallskip\vskip .1in}}
\def\ref{\noindent \parshape2 0truein 4.25truein 0.25truein 4truein}
\def\caption{\noindent \captiontype
             \parshape=2 0truein 4.25truein .125truein 4.125truein}

\def\footnote#1{\edef\fspafac{\spacefactor\the\spacefactor}#1\fspafac
      \insert\footins\bgroup\footnotetype
      \interlinepenalty100 \let\par=\endgraf
        \leftskip=0pt \rightskip=0pt
        \splittopskip=10pt plus 1pt minus 1pt \floatingpenalty=20000
        \textindent{#1}\bgroup\strut\aftergroup\strut\egroup\let\next}
\skip\footins=12pt plus 2pt minus 4pt 
\dimen\footins=30pc 

%
%

\def\@{\spacefactor 1000}

\def\,{\pcomma} 
\def\pcomma{\relax\ifmmode\mskip\thinmuskip\else\thinspace\fi}

\def\oversim#1#2{\lower0.5ex\vbox{\baselineskip=0pt\lineskip=0.2ex
     \ialign{$\mathsurround=0pt #1\hfil##\hfil$\crcr#2\crcr\sim\crcr}}}

\pageno=1        
\input epsf

\def\runningtitletext{DUST IN PROTOPLANETARY DISKS}
\def\runningauthortext{BECKWITH, HENNING, \& NAKAGAWA}

\null
\firstpageskip


{\baselineskip=14pt
\title{Dust Properties and Assembly of Large Particles}
\title{in Protoplanetary Disks}
 }

\vskip .3truein
\name{Steven V. W. Beckwith}
\affiliation{Max Planck Institute for Astronomy, Heidelberg}
\affiliation{Space Telescope Science Institute}
\vskip .2truein
\name{Thomas Henning}
\affiliation{Astrophysical Institute and University Observatory, Jena}
\vskip .1truein
\leftline{and}
\vskip .1truein
\name{Yoshitsugu Nakagawa}
\affiliation{Kobe University}
\vskip .3truein

\abstract{Recent research on the buildup of rocks from small dust grains has reaffirmed that grain growth in protoplanetary disks should occur quickly.  Calculation of growth rates have been made for a variety of growth processes and generally predict high probabilities of sticking in low-velocity collisions that may be brought about in a number of ways in protoplanetary disks.  Laboratory experiments have measured sticking coefficients for some materials largely confirming the calculations.  Although the detailed velocity fields of disks are not well understood, many of the important processes leading to particle collisions and grain growth have been studied theoretically and demonstrate likely paths by which dust is assembled into planets.  Calculations of the radiative properties of particles with various size distributions show that large particles should produce observable changes in the spectral energy distributions of disks.  Changes of the sort predicted are, in fact, observed, but their interpretation is ambiguous; there are other ways to produce the observed changes that do not require grain growth, so the evidence is currently inconclusive.  The major uncertainties can be overcome with the next generation of millimeter-wave interferometers, and it seems likely that a firm case for grain growth could be established within a decade.}

\voffset 2truecm

\mainsection{{I}.~~Introduction}
Planets are built from tiny interstellar dust grains by the steady accumulation of the smaller particles, at first through sticking but later aided by gravity between larger planetesimals, and subsequently by the accretion of gas onto the rocky cores.  Such growth should occur naturally in the circumstellar disks created when young stars are born.  The disks provide protected "wombs" for the development of the large bodies during the course of a few million years or more.  Although disks are not the only places where planets might have grown, they are theoretically well suited for planet formation and observed to be ubiquitous, so that they are the clear favorite for planetary birthsites at the time of this review.  A demonstration of the growth of dust into large bodies is the one open issue that remains to demonstrate that circumstellar disks - hundreds of examples of which are routinely studied - do spawn planetary systems.  Such a demonstration would also imply that planetary systems are common features of nearby stars, just as disks are commonly found around young stellar objects.

In the earliest stages, particles are built up by coagulation and sticking as they bump into one another in the disks.  In about $10^4$ yr, particles can grow from micron-size to meter-size through a variety of collisional processes: Brownian motion, settling, turbulence,  and radial migration, the dominant mode depending on physical conditions within the disk.  After another $10^4$ years or so, the small bodies can attain sizes of order kilometers.  Beyond 1 km, gravitational attraction causes these planetesimals to grow by pairwise collisions and later by runaway growth until some planets will be present after ~1 Myr (Weidenschilling 1988; Weidenschilling \& Cuzzi 1993; Lissauer 1993).  Accretion of gas should create gaps in the disks that may present a natural limit to the growth of planets over very long times.  Figure 1 shows the three stages schematically.  The timescales are uncertain but still well enough constrained to be shorter than the observed lifetimes of disks around young stars (e.g. Strom et al. 1993).  If the theory is even approximately correct, there should be large bodies within currently observed circumstellar disks.

The buildup of a small, rocky planet is longer than the lifetime of any observer by several orders of magnitude, so it is not possible to watch this evolution.  Instead, we must be content to compare the particle properties of disks of different ages and infer the growth of large bodies from trends in the size distribution with age.  Even the indirect detection of planetary-mass bodies in circumstellar disks is difficult when the disk mass is greater than the planetary mass, but there is hope that the indirect effects of larger bodies will be observable in a decade or sooner.

The theory of grain coagulation delineates the time scales over which to study disks and provides some limits on what is plausibly observed.  The theory is checked by laboratory experiments for small scale growth, where major bottlenecks might occur.  It is possible to create tiny grains similar to interstellar dust in the laboratory under physical regimes like those thought to occur in disks.  The growth rates of these particles can be checked against the theory to ensure that the physics of grain growth is well understood.  Experimentally measured growth rates for small particles largely verify the calculations and lead us to believe that the growth of the smallest particles is reasonably well understood.  It is  difficult to verify the growth of larger bodies.  It will, therefore, be important to detect some large bodies near other stars to check the calculations of early planet evolution.

\vskip .1in
{
\centerline{\epsfxsize=4.25in \epsfbox{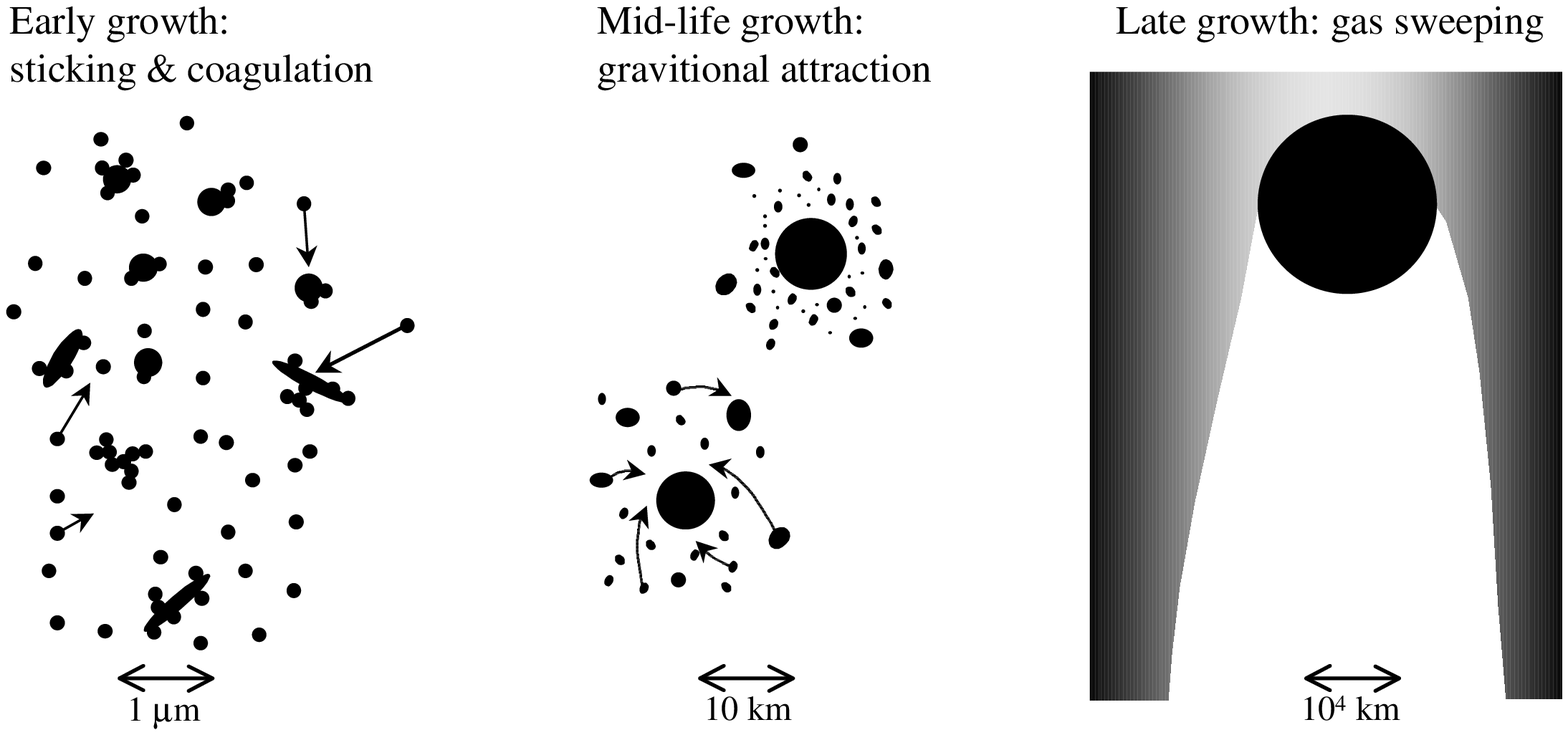}}
\vskip .1in
\caption{Figure 1.\capskip This figure depicts the sequence of events leading to planet formation: grain growth through particle-particle interactions, growth of planetesimals by two-body interactions aided by gravity, and the accretion of gas via gravitational attraction.}
}
\vskip 0.15 in

There are enough known disks around nearby young stars to provide good samples for detection of grain growth.  There are now methods for discerning the growth of dust grains to pebbles and rocks (Miyake \& Nakagawa 1993) via the submillimeter opacities.  These methods have had limited application to date but give interesting hints of grain growth.

This review discusses several aspects of this problem, including the physics of the growth process, the observable effects on radiation from collections of the particles, and the extant observations designed to see evidence of grain growth.  There has been progress on all these fronts since the last Protostars and Planets meeting in 1990, spurred by an increasing interest in extra-solar planet formation that is driven by detections of extra-solar planets.  Most of this progress has been in laboratory work and new calculations of the properties of small particles as they pertain to coagulation and growth.  There were already tantalizing hints at the time of the last conference, {\it Protostars \& Planets III} that particle growth had been detected in circumstellar disks; a few new observations attempt to clarify these hints. We will concentrate on progress made since PPIII.

\mainsection{{I}{I}.~~Grain evolution in disks}
\nobreak
\backup
\subsection{A.~~Structure of protoplanetary disks}
The idealized infinitely thin, continuous disk assumed in early theories (e.g. Lynden-Bell \& Pringle 1974; Pringle 1981; Adams \& Shu 1986) has given way to a disk with vertical and radial structure and with gaps or holes needed to account for various observations (Bodenheimer 1995; Papaloizou \& Lin 1995).  Stellar radiation provides at least minimal heating that is often enhanced by accretion energy as material from the disk falls onto the star.  Disks heated mainly by accretion are referred to as {\it active}, whereas those heated entirely by the stars are called {\it passive}.  The spectral energy distributions of most simple disks can be fitted reasonably well assuming they are flared (Kenyon \& Hartmann 1987), centrally heated from radiation near the star (Chiang \& Goldreich 1997) and have radially decreasing temperatures (Adams, Lada, and Shu 1988).  There are striking exceptions to this picture -- HL Tau, for example, and almost all of the ''flat-spectrum'' sources; these presumably have a number of different components -- dense, disk-like regions, outflow cavities, the surrounding molecular cloud cores -- that contribute to the SEDs and complicate the analysis (Calvet et al. 1994)

Disks usually have outer radii between a few and a few hundred AU (Appenzeller et al. 1984; Edwards et al. 1987; Lay et al. 1997) often with sharp outer boundaries (O'Dell \& Wenn 1994; McCaughrean \& O'Dell 1996).  Young disks are thought to be optically thick at all wavelengths shortward of ~100\,$\mu$m.  Typical disk masses are between ~0.001 and 0.1\,M$_\odot$, although there is some uncertainty about this range (Adams et al. 1990; Beckwith et al. 1990; Henning et al. 1993; Andr\'e \& Montmerle 1994; Osterloh \& Beckwith 1996).  These properties accord well with those assumed for the disk in the early solar system from which our planets were born (Safronov 1969; Hayashi 1981; Cameron 1988).

Disk structure is important to the study of particle growth in two ways.  First, the presence of convection, turbulence eddies, and other velocity gradients can strongly affect the growth of particles, since the growth rates depend on the relative particle velocities.  Second, observations of disks must either take disk structure into account for their interpretation or else be designed to be relatively unaffected by the structure.  These effects are discussed in the sections to follow.  There is enough freedom in models of disks to produce many kinds of local structure, and important effects such as turbulence eddies, sedimentation, and radial migration of matter are currently beyond observation.  They are also a matter of debate among theorists.  While several instability mechanisms might in principle maintain nebular turbulence, the presence of turbulence and the dominance of any energy source has yet to be demonstrated definitively (Cabot et al. 1987; Balbus \& Hawley 1991a, b; Dubrulle 1992, 1993; Tscharnuter \& Boss 1993; Nakamoto \& Nakagawa 1994, 1995; Kley et al. 1993; Stone \& Balbus 1996; Klahr et al. 1998).  As a result, there is considerable uncertainty about the actual evolution of particles in disk theories, and a wide range of physical conditions needs to be considered in theories and laboratory studies of grain growth relevant to planet formation.

\subsection{B.~~Physical processes leading to grain growth}
In the outer regions of an accretion disk, most of the interstellar dust remains unaltered during  passage through the weak shock front when the disk is first created. Only volatile ice mantles around the refractory grain cores partly sublimate, but water ice efficiently forms when the water vapor cools behind the shock (Lunine et al. 1991, Neufeld and Hollenbach 1994). Pollack et al. (1994) summarize the expected grain components, their optical properties, and sublimation temperatures immediately after disk formation. The main species are: (1) olivines (Mg$_x$Fe$_{2-x}$SiO$_4$ with x=1.4), (2) orthopyroxenes (Mg$_x$Fe$_{1-x}$SiO$_3$ with x=0.7), (3) quartz (SiO$_2$), (4) metallic iron (Fe), (5) troilite (FeS), (6) refractory and volatile organics, and (7) water ice.  Henning and Stognienko (1996) update these calculations by taking into account more recent determinations of optical constants and a possible fluffy structure of the grains.  They also found that the grain opacities depend on the fraction of iron that is in iron grains or silicates.  

If the grains are transported inwards during the viscous evolution, the higher temperatures will anneal and eventually destroy them (see, e.g., Finocchi et al. 1997, Gail 1998).  Gail (1998) assumes chemical equilibrium and concludes that crystalline silicates, iron, and aluminium compounds are the most important grain components in the warmer parts of the disk. Such equilibrium condensation calculations  have been extensively performed (see, e.g., Grossman 1972, Cameron and Fegley 1982, Yoneda and Grossman 1995). The results are uncertain, however, because a good  understanding of the chemical processes during grain condensation, growth, and destruction requires a kinetic approach which cannot yet be done owing to unknown reaction rates for oxygen-rich chemistry (see also Prinn 1993).

The condensation of ice onto refractory cores in the outer regions of the disk is important for the growth of submicron-sized grains but is irrelevant to growth from micron-sized grains to meter- and kilometer-sized planetesimals (e.g. Preibisch et al. 1993).  But the surface structure of the dust grains may be considerably modified by repeated condensation and sublimation caused by transport processes or heating events. Such ``flares'' were also discussed as a possible mechanism for the production of chondrules found in meteorites (see Hewins et al. 1996).

For much of the last 30 years, planetesimal formation was thought to be the result of a gravitational instability in a dense dust layer leading to fragmentation into kilometer-sized bodies (Safronov 1969, Goldreich and Ward 1973).  Weidenschilling (1980) was the first who noted the difficulty with the particle layer instability.  Cuzzi et al. (1993) and Weidenschilling (1995) demonstrated that such an instability is unlikely to occur.  Particles must grow to at least meter size before they can decouple from shear-induced turbulence and slow to the low velocities required for the instability to occur. The radial velocity dispersion induced by drag will further delay the onset of instability until the mean size is in the range of 10-100 m.  Thus, the formation of planetesimals has to be explained by the collisional aggregation of particles.   

It is difficult to calculate this process from first principles,  because there are complex feedback mechanisms in the two-fluid system whose relative importance is not well known: aerodynamical drag, thermal structure coupled to changing dust opacities, charge state of the medium and coupling to magnetic fields, for example.  It is imperative to understand particle growth from collisions between particles, however, if we want to estimate the likelihood of
creating planetesimals and planets.

Particles collide because they are moving relative to one another. The sources of relative velocities in laminar nebulae  are Brownian motion, $v_{th}$, sedimentation, $v_s$, and  radial drift, $v_r$. In turbulent systems, turbulence-driven motion, $v_{tur}$, can additionally be of importance for grains/bodies of certain sizes. The equations for the relative velocities are:

$$ \Delta v_{th} = \sqrt{8 kT/\pi \mu} \quad \hbox{with} \quad \mu = (m_1 m_2)/(m_1 + m_2) ,  \eqno(1) $$

$$ \Delta v_s =  {\Omega}^2 z \quad \vert \tau_{f1} -  \tau_{f2}\vert , \eqno(2) $$

$$ \Delta  v_r =  2 \Omega \triangle V \quad \vert \tau_{f1} -  \tau_{f2}\vert , \eqno(3) $$

$$ \Delta  v_{tur} \simeq  u_s/t_s \quad \vert \tau_{f1} -  \tau_{f2}\vert  \quad \hbox{for} \quad \tau_{f1}, \tau_{f2} < 
t_s , \eqno(4) $$

$$ <\Delta  v_{tur}> =  1.33 u_l  \quad \sqrt{\tau_f/t_l}  \quad \hbox{for} \quad t_l > \tau_{f1}, \tau_{f2} > t_s , 
\eqno(5) $$

\noindent  (Weidenschilling 1977, V\"olk et al. 1980, Mizuno et al. 1988).  Here, $\Omega$ is the Keplerian angular frequency; $T$ the temperature; $m_1$ and $m_2$ the particle masses; $\tau_{f1}$ and $\tau_{f2}$ the friction times (see IIIA for a more detailed discussion), $\tau_f$ = max($\tau_{f1}$, $\tau_{f2}$); $z$ the vertical distance from the midplane; $\triangle V$ the velocity difference between Keplerian velocity and rotational gas velocity; $u_s$ and $t_s$ the velocity and turn-over time of the smallest turbulent eddy; $u_l$ and $t_l$ the velocity and turn-over time of the largest turbulent eddy. In the case of turbulence-driven motion, we have distinguished between two cases of the coupling of the grains to the gas ($<\Delta v_{tur}>$ is the rms value of the relative velocities).

A comparison of equations (1) through (5) shows that the particles with equal friction times must couple to the turbulence (second case, eq. 5) to obtain substantial relative velocities; otherwise, only the Brownian motion is important.  The collisions caused by Brownian motion are relevant only for small (about micron-sized) grains because of the inverse mass dependence. On the other hand,  an efficient coupling to the turbulence in a disk with $\alpha$=0.01, say, is possible for mm-sized grains to m-sized bodies, calculated for the disk model of Bell et al. (1997) at 10 AU radius and considering the friction time for compact particles. Grains with very small friction times essentially comove with the gas. One needs friction times larger than the turn-over time of the
smallest eddies for the start of an efficient growth by turbulence. In a laminar nebula, a  dispersion in friction times is always required for an efficient growth. 

\subsection{C.~~Homogeneous growth regime}
Most models for collisional grain growth in protoplanetary accretion disks assume a locally homogeneous distribution of particles. There are classically two types of models: radial one-dimensional (1D) models that average the density distribution vertically through the disk (see, e.g., Schmitt et al. 1997), and vertical 1D models for the coagulation of particles by settling to the midplane at a given radius (see, e.g., Weidenschilling and Cuzzi 1993). 

Several authors have studied the 1D radial case each incorporating more detail into the calculations (see, e.g., Morfill 1988; Mizuno et al. 1988; Mizuno 1989). Ruden and Pollack (1991) treated the dynamical evolution of a protoplanetary disk under the hypothesis that the only source for the turbulent viscosity is thermal convection. They discussed grain settling and coagulation in the outer solar system when turbulence ceases. Schmitt et al. (1997) coupled the disk and dust evolution. They distinguished three phases during the first 100 years of disk evolution: an early phase in which the smallest particles disappeared quickly due to Brownian-motion driven coagulation and produced a relatively narrow mass distribution; an intermediate phase of self-similar growth due to turbulence where the size distribution can be described by scaling laws; and a late phase where the most massive particles decouple from the gas and drift motions become an important source of relative velocities. Stepinski and Valageas (1996) consider the global evolution of single-sized non-coagulating particles in radial models with $\alpha$ between 0.01 and 0.001. They conclude that particles larger than 0.1 cm are not really entrained in the gas. They also studied the formation and radial distribution of km-sized icy planetesimals assuming that the size distribution is always narrowly peaked around a mean value for a given radial location and time (Stepinski and Valageas 1997).

A shortcoming of radial models is the inability to treat the formation of a dust subdisk due to settling. Weidenschilling (1980), Nakagawa et al. (1981, 1986), and Weiden\-schil\-ling and Cuzzi (1993) developed numerical 1D models for settling with coagulation. The main steps of the evolution are: (1) Micron-sized grains uniformly suspended in the gas grow by Brownian motion, (2) grain aggregates begin to settle and sweep up smaller ones in a kind of runaway growth, and (3) the aggregates grow to centimeter-sizes and create a dense dust layer (Weidenschilling 1997). Weidenschilling (1997) treatede the growth from microscopic grains to cometesimals at 30 AU in detail. The dependence of drag-induced velocities on particle size produced a stage in which most of the mass is concentrated in a narrow size range between 10 and 100 m.  Dubrulle et al. (1995) found that the turbulence does not influence the timescale of sedimentation, but merely determines the equilibrium scale height of the dust subdisk. The smallest grains that can experience some kind of sedimentation in the presence of turbulence have sizes of the order of 150$\alpha$\,cm, resulting in a size limit of about 1\,cm for $\alpha$=0.01. Sekiya (1998) showed that for mm-sized aggregates even very weak turbulence stirs up the dust aggregates.  This calculation considered only locally produced turbulence corresponding to the calculations by Cuzzi et al. (1993) and Champney et al. (1995), not an $\alpha$-disk.

Radial transport becomes important over the timescale of settling. The results of the 1D investigations show that consistent 2D/3D models or at least 1+1D models are necessary for an adequate description of the dust dynamics.

\subsection{D.~~Inhomogeneous growth regime}
In the classical models, turbulence is treated only statistically as a source of random velocities important for diffusion and the calculation of turbulence-induced relative velocities.  These models assume that particles remain uniformly distributed in a turbulent velocity field.  This prediction is only valid if the typical length scales are larger than the eddy scales.

Squires and Eaton (1991) were the first to show that turbulence can lead to particle concentration. Cuzzi et al. (1996) calculated this particle concentration for a protoplanetary disk with weak Kolmogorov-type tubulence.  Concentrations of $10^5$ to $10^7$ occur for particles with radii between 0.1 and 1 mm in convergence zones between eddies.  The particle trapping mechanism may have dramatic consequences for the formation of chondrules due to its size-selective nature (see, Cuzzi et al. 1996).

Klahr and Henning (1997) showed that in certain vortical flows particles are trapped and concentrated in the gas flow.  Slowly rotating eddies and a gradient of the vertical component of the gravitational force produce this trapping.  The gradient causes particles in the upper half of an eddy to fall faster than in the lower half.  If this concentration is stronger than the dispersion of grains due to centrifugal forces, the particles are concentrated in the interior of the eddies.  The strongest effect was observed for millimeter-sized particles; they can be concentrated by a factor of 100 within 100 years.Three-dimensional calculations by Klahr et al. (1998) indicate that the particle concentration mechanism is not just the result of 2D approximations.  Figure 2 shows the results of a calculation in which an initially homogeneous distribution becomes inhomogeneous after 160 years. However, we should note that the possibility to concentrate or to disperse small dust particles depends on the structure of the flow pattern (3D disk turbulence vs. slow and long-lived vortical circulation) and the stability of the vortices (Adams and Watkins 1995; Klahr and Henning 1997; Klahr et al. 1998;  Brandenburg and Hodgson 1998).

\vskip 0.1in
{
\centerline{\epsfxsize=4.25in \epsfbox{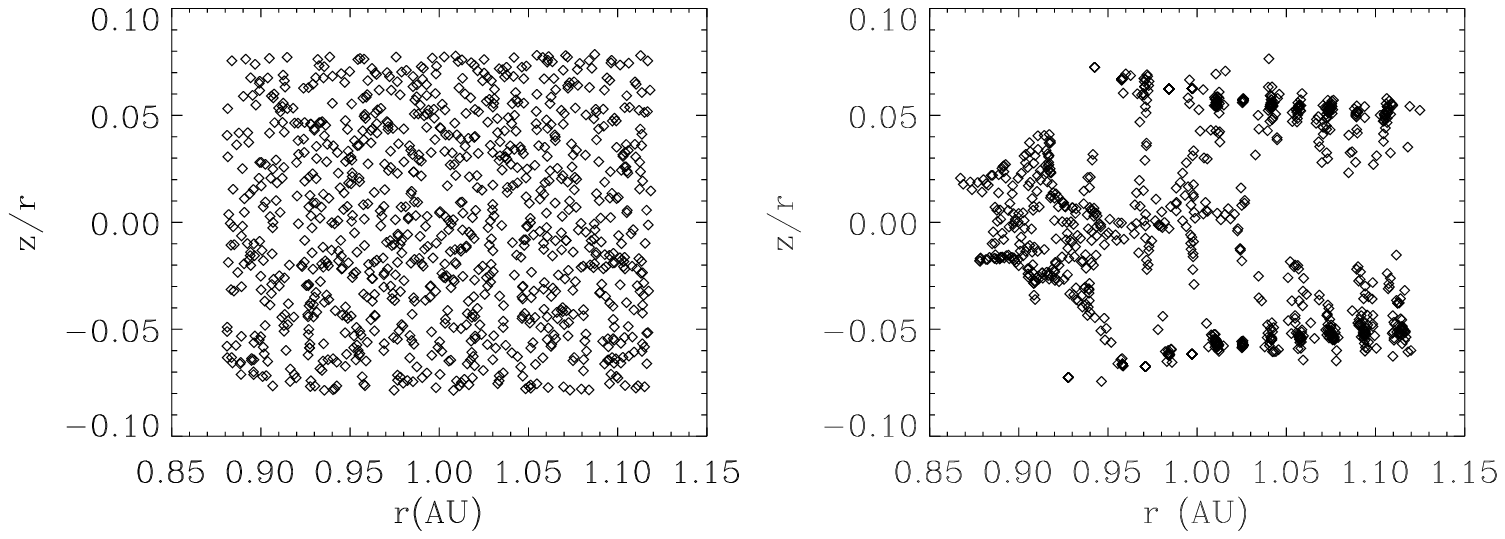}}
\vskip 0.1in
\caption{Figure 2.\capskip The left hand portion shows an initially homogeneous distribution of grains that becomes quite inhomogeneous after 160 years in the right hand portion because of trapping in eddies (Klahr et al. 1998).}
}
\vskip 0.15 in

\noindent Tanga et al. (1996; see also the earlier related study by Barge and Sommeria 1995) considered one more possibility of particle concentration via vorticity. They introduced an ad hoc system of regular vortices driven by differential rotation of the disk with rotational axes parallel to its angular momentum. In this case, the concentration mechanism is based on Coriolis forces and much larger particles would be concentrated compared with the cases discussed before.

\mainsection{{I}{I}{I}.~~How well is particle growth understood?}
\backup
\subsection{A.~~Aerodynamical behaviour of particles in a dilute gas}
The interaction of dust particles and small bodies with sizes below 1\,km is determined by their aerodynamical coupling to the dilute gas of the protoplanetary accretion disk and not by gravitational forces. This coupling can be characterized by a friction (or response) time $\tau_f$ which for dust particles and aggregates is much smaller than the orbital period; i.e. the particles are strongly coupled to the gas. For the conditions in protoplanetary disks, the coupling of submicron- to centimeter-sized particles to the gas is  dominated by collisions of grains with individual molecules. In this regime, the molecular mean free path, $l_{mfp}$, is larger than the characteristic particle size, $a$.  The friction time of a spherical dust particle of mass, $m$, and geometrical cross section, $\sigma_g$, is given by:

$$ \tau_f = {m \over 4/3 (\delta P) \sigma_g \rho_g v_g} \eqno(6) $$

\noindent Here $\delta P$ is the momentum transfer efficiency factor and the quantities $\rho_g$ and $v_g$ denote the gas density and the mean thermal velocity of the molecules, respectively. The material parameter $\delta P$ is always close to 1.  Experiments by Hutchins et al. (1995) and Blum et al. (1996) verified the validity of (6) and showed that a single value $\delta$$P$= 1.11 $\pm$ 0.17 describes both spherical SiO$_2$ grains and small aggregates. These experiments confirmed the results of numerical Monte-Carlo simulations on the aerodynamic properties of fractal aggregates (Meakin et al. 1989).

The strength of gas-grain coupling depends on the ratio of mass to cross section, so $\tau_f$ depends on whether the particles are compact - spheroids, say - or fluffy, open structures (Ossenkopf 1993). If the aggregates are fractals, we can introduce two fractal dimensions $D_m$ and $D_\sigma$:

$$ {m \propto R^{D_m}} \quad \hbox{and} \quad  {\sigma_g \propto R^{D_\sigma}} \eqno(7) $$

\noindent and show that $\tau_f$ $\propto$ R$^{(D_m - D_{\sigma})}$. In 3-dimensional systems, $ 1 \leq D_{\sigma} \leq D_m \leq 3 $.  Note that $D_m = 3$ and $D_\sigma = 2$ for compact particles, so that $\tau_f \propto R$.

\subsection{B.~~Aggregation process}
Ballistic cluster-cluster aggregation, de\-noted BCCA, is the process in which clusters that collide stick together without restructuring and with sticking probability 1. This process leads to the production of open, fluffy aggregates.  Numerical simulations of BCCA yield $D_m$ values between 1.81 and 2.00 depending on the numerical algorithm and the selection of collision partners (Meakin 1984).  Changing the sticking probability does not dramatically change $D_m$.  Collisions of polydisperse clusters always result in somewhat higher fractal dimensions than collisions between monodisperse clusters (Brown and Ball 1985).  Ballistic particle-cluster aggregation, BPCA, in which clusters grow by the addition of individual particles, give values of $D_m$ close to 3.0. The dimension $D_{\sigma}$ must always be smaller than $D_m$ and is often assumed to be close to 2. This means that BCCA clusters have friction times that are nearly independent of radius. For such aggregates and typical conditions in the solar nebula at 1\,AU, the relative velocities are between $10^{-3}$\,m\,s$^{-1}$ - Brownian motion - and $10^{-2}$ m s$^{-1}$ - sedimentation.  For turbulence-driven BPCA growth, the mean mass of the mass distribution scales with the 
evolutionary time, $t$, as $t^6$, whereas the scaling for BCCA particles is exponential with time (Mizuno et al. 1988).

Numerical simulations of the growth process driven by Brownian motion under conditions  appropriate for protoplanetary accretion disks (Kempf et al. 1998) show only a small dispersion of the friction times leading to small relative velocities and slow growth.  The numerical calculations also demonstrated that the friction time increases rather slowly with radius: $\tau \propto R^{0.2}$; $D_m \sim 1.8, D_{\sigma} \sim 1.6$. No such simulations exist for turbulence-driven growth or coagulation driven by drift velocities.  Experiments on the aggregation of micron-sized dust grains in a turbulent gas (collisional velocities in the experiment $\sim 20$\,cm\,s$^{-1}$) and during sedimentation (collisional velocities $\sim 1$\,cm\,s$^{-1}$) confirm the formation of fractal dust grains with $D_m$ between 1.9 and 1.7 in the two regimes (Blum et al. 1998 and references therein).  The slow increase of $\tau_f$ means grain growth is also slow. 

The friction time, $\tau_f$, must increase markedly  ($D_m > 2$) to produce efficient, turbulence-driven coagulation ($\tau_f > t_s $) (see equation (5)).  Rapid growth will also occur when there is a reservoir of small particles (e.g. particles accreted from the molecular cloud core) leading to a BPCA-type growth process or during ``runaway'' growth where very large particles (decoupled from the mass distribution) sweep up a large amount of small grains (Safronov 1972; Weidenschilling and Cuzzi 1993; Schmitt et al. 1997;  Weidenschilling 1997). In this case, the statistical coagulation equation cannot be used any longer (see, e.g., Tanaka and Nakazawa 1994). 

Experimental results  suggest the possibility of enhanced coagulation rates of small iron particles by the presence of a magnetic field (Nuth et al. 1994;  Nuth and Wilkinson 1995). Whenever a magnetic field of about 100 Gauss was present, the iron particles coagulated very rapidly and formed complicated aggregates. The relevance of this process for protoplanetary accretion disks depends on the possibility of the magnetization of small iron grains. 

\subsection{C.~~Critical velocities for grain sticking}
The growth process is governed by the sticking probability, $S$, and the strength of the adhesion forces holding the aggregates together.  It is generally assumed that van der Waals forces are responsible for the sticking of small particles.  The sticking probability for two colliding particles depends on the collision velocity, masses, shapes, and material properties.  There may also be a critical velocity for sticking, $v_{cr}$.  The existence of a critical velocity implies there is a steep transition from the velocity range where sticking occurs ($v < v_{cr}$, $S \sim 1$) to the velocity range where the particles bounce ($v > v_{cr}$, $S \sim 0$). Even for the simplest case of spherical particles, the calculated and measured values of $v_{cr}$ disagree.  There are many complicating factors to be understood in these collisions, so this disagreement is not surprising.

Chokshi et al. (1993; see also Dominik and Tielens 1997) modelled the dynamics of sticking collisions between dust particles.  Two colliding particles form a contact ``neck'', and the released energy due to the decrease of surface energy accelerates the particles towards each other. Elastic compression decelerates the particles, reverses the motion, and finally leads to acceleration in the opposite direction. When the grains again reach the point of first contact, they are still bound by a contact neck. After further separation, the binding neck becomes unstable and the contact can rupture.  Whether or not rupture occurs depends on the amount of kinetic energy dissipated during the collision.  In the model developed by Chokshi et al. (1993), energy dissipation is assumed to be dominated by elastic surface waves. The critical velocity they derived for central collisions between two spherical grains is:

$$ {v_{cr} = 1.07 { \gamma^{5/6}  \over  E^{1/3} r^{5/6} \rho^{1/2}}}, \eqno(8) $$

\noindent where $\gamma$ is the surface energy per unit area, $r$ is the reduced radius ($r=$ ${r_1 r_2}\over {r_1 + r_2}$, $r_1$ and $r_2$ are the radii of the colliding particles), and $\rho$ is the mass density of the spheres. $E$ is a material quantity which depends on the Poissons's ratios and Young's moduli and reflects the elastic properties.\footnote{\dag}{The factor in (8) has to be 1.07 instead of the 3.86 erroneously given in Chokshi et al., and the functional dependence has to be E$^{1/3}$ instead of E as misprinted in Dominik and Tielens.} A compilation of material properties for astrophysically relevant sytems is given by Dominik and Tielens (1997).  For micron-sized quartz spheres, the critical velocity is calculated to be $\sim 5$\,cm\,s$^{-1}$. 

Poppe et al. (1997, 1998) imaged the trajectories of micron- and sub-micron-sized particles colliding with a target to determine the sticking characteristics for different materials.  For spherical silica particles  with smooth surfaces and  with $r = 0.6\,\mu$m, the critical velocity is 1.2\,m\,s$^{-1}$ which is an order of magnitude higher than predicted by equation (8).   The  shape of the particles is an important property for sticking.  The sticking probability increases significantly for irregularly shaped dust grains especially for submicron grains.  In some experiments, no critical velocity is seen, and a significant fraction of irregular particles stick even at impact velocities exceeding 20 m s$^{-1}$. 

Measurements with an atomic force microscope demonstrate that the pull-off forces are in agreement with the theoretical predictions. The discrepancy between experimental and theoretical values for $v_{cr}$ in the case of the spherical particles is probably caused by an underestimate of the actual energy dissipation rate in the theoretical calculations. The general conclusion from the experiments is that micron- and sub-micron-sized grains stick with high probability as long as the relative velocities are below a few m\,s$^{-1}$ - a limit which should not be reached by such particles in protoplanetary disks with $\alpha$ values below 10$^{-1}$. 

An interesting discovery of the experimental work is that collisional charging occurs. The charge transfer in a collison between micrometer-sized silicon particles and a flat quartz target surface results in averaged values of separated elementary charges per impact energy of 10$^{-5}$ to 10$^{-4}$ C J$^{-1}$. The charging is approximately proportional to the collision energy. Therefore, electrostatic forces between charged grains may be important for the growth process, but the sign and magnitude of the charges are difficult to predict a priori.

Bridges et al. (1996) and Supulver et al. (1997) addressed the further growth from centimer-sized particles to meter-sized bodies. They found that several types of water frost-coated surfaces stick together when brought into contact at low enough sticking velocity, $< 0.4$\,cm\,s$^{-1}$. Their experiments show that the structure of frost is important for sticking.  They concluded that the formation of surface layers of frost is a necessary step for providing a sticking mechanism for the growth of larger particles in low-temperature regions of protoplanetary accretion disks. It might be interesting to note that the formation of frosty surfaces plausibly results from sweepup of tiny grains onto the surfaces of the larger particles in relative motion. The layer would be regenerated following compacting bounces.

If larger particles (cm- to meter-sized) reach relative velocities of the  order 10\,m\,s$^{-1}$ or more, the collisons may cause compaction and disruption.  High velocities for the larger particles would lead to fragmentation and the production of a reservoir of small particles.

\subsection{D.~~Restructuring, compaction, and disruption in aggregate-aggre\-gate collisions}
Aggregates with open BCCA-type structures cannot grow indefinitely through collisions. Collisions with larger grains or other aggregates should lead to restructuring and compaction at a certain mass and velocity threshold. Compaction is necessary to increase the fractal dimension ($D_m > 2$) which increases the frictional coupling and leads to a  more efficient growth regime.  This was the motivation for Weidenschilling (see, e.g. Weidenschilling and Cuzzi 1993) to start with a fractal dimension of 2.11 in the size range 1\,$\mu$m to 0.1\,mm and then to introduce compaction starting with grain sizes of 1\,cm. Although such a compaction is likely, the values of the size limits are unknown, and the change from $D_m = 2.11$ to 3 is arbitrary.  And compaction may not occur only in collisions; recurrent heating events could also create compact particles.

Theoretical calculations of the compaction by collisons between aggregates by Dominik and Tielens (1997) predict a sequence of events depending on the ratio between impact energies and critical energies for the different processes: (i) sticking without restructuring, (ii) losing monomers, (iii) maximum compression, and (iv) catastrophic disruption. Their evaluation of the different processes showed that rolling is the most efficient mechanism for restructuring.

For velocities below 1\,cm\,s$^{-1}$, the sticking probabilities in collisions of spherical SiO$_2$ aggregates are close to unity (Wurm and Blum 1998). There is no restructuring in turbulence-driven growth until collisional velocities of 20\,cm\,s$^{-1}$.  Measurements of the rolling friction between SiO$_2$ aggregates and a massive target by Wurm (1997) give a lower limit of $5.0\times 10^{-10}$N, about a factor of 5 larger than the value used by Dominik and Tielens (1997).  Bouncing and fragmentation dominated the results of collisions down to $\sim 1$\,m\,s$^{-1}$. Between 0.1 and 1\,m\,s$^{-1}$ (and especially close to the critical velocity) restructuring occured.   

The sequence found in the experiments is in qualitative agreement with the theoretical predictions of Dominik and Tielens (1997), but the threshold velocities for the different processes are higher by at least a factor of 2 than the calculated values, mainly because the calculations underestimate the frictional forces. These results imply that at least in the beginning of the sedimentation-driven growth, no restructuring will occur and the aggregates have to reach centimeter sizes before compaction occurs.

\mainsection{{I}{V}.~~Observational consequences of particle size}

\subsection{A.~~Theoretical changes of particle opacity}
As dust particles grow by collisions, the particle mass opacity, $\kappa_\nu$ (cm$^2$\,g$^{-1}$), changes.  Changes in opacity can in principle be observed in circumstellar disks, thus providing the signature needed to demonstrate that disks breed large particles and planets.  The opacity at any wavelength, $\lambda$, will not change until a significant fraction of the particle mass is in particles with sizes of order $\lambda$ or greater.  

The flux density, $F_\nu$, from a disk viewed face on at distance, $D$, is:

$$ {F_\nu = {1\over D^2} \int_{r_0}^{R_D} B_\nu (T(r)) (1 - e^{-\tau_\nu (r)}) 2 \pi r dr}, \eqno(9) $$

\noindent where the interior and exterior disk radii are $r_0$ and $R_D$, $B_\nu$ is the Planck function, $T$ is the temperature, $\tau_\nu$ is the optical depth, and $r$ is the radial variable (cf Beckwith et al. 1990).  The optical depth can be written in terms of the surface density, $\Sigma$, and mass opacity coefficient, $\kappa_\nu$ as: $\tau_\nu (r) = \kappa_\nu \Sigma (r)$.  If the particles are small and the wavelengths are much larger than the size of the largest particles, the disk should be transparent, $\tau_\nu < 1$, and the emission is in the Rayleigh-Jeans regime, $B_\nu \approx 2 k T \nu^2$.  In this limit:

$$ {F_\nu \approx \kappa_\nu \nu^2 {4 \pi k \over D^2} \int_{r_0}^{R_D} T(r) \Sigma(r)r dr}. \eqno(10) $$

\noindent Regardless of the disk structure, the flux density is directly proportional to the particle mass opacity, $\kappa_\nu$.  Although the absolute value of $\kappa_\nu$ is difficult to determine without a knowledge of $T(r)$, $\Sigma(r)$, and $R_D$ ($r_0$ is usually not relevant in this limit), the frequency dependence can be observed directly.

Particle emission in most disks becomes optically thin at millimeter wavelengths.  The emission emerges from the outer parts of disks where $T$ is a few tens of Kelvin, and $\Sigma$ should be small, so that the approximations made in Eqn. 10 are valid (Beckwith et al. 1990).  Observations of disks over a range of wavelengths, between 0.6 and 2.7\,mm, say, should provide a direct measure of the emissivity exponent, $\beta$.  If a mixture of particle types and sizes is present, the observed value is an average over the different constitutents.  

At long wavelengths, $\lambda > 0.1$\,mm, $\kappa_\nu$ is expected to scale as a power of the frequency: $\kappa_\nu \propto \nu^\beta$.  For compact spherical particles smaller than the observing wavelength, $\beta = 2$ for metals and insulators under a wide range of conditions (Bohren and Huffman 1983; Emerson 1988).  The power law exponent, $\beta$, can be as small as 1 for certain types of materials - amorphous carbonaceous material, for example - and even smaller over limited ranges, but it must go to 2 at long enough wavelengths to fulfill causality relations, and it is expected that $\beta \approx 2$ at millimeter wavelengths for interstellar dust particles.  Careful attempts to measure both $\kappa_\nu$ and $\beta$ in interstellar clouds have, indeed, yielded $\beta \approx 2$ and values of $\kappa_\nu =(0.002--0.004)(\lambda /1.3{\rm mm})^{-2}$\,cm$^2$\,g$^{-1}$ (Hildebrand 1983; Draine and Lee 1984).  

Rocks, asteroids and planets are opaque to radiation at $\lambda \sim 1$\,mm, in which case $\beta = 0$.  Pebbles with sizes of order 1\,mm should have exponents with intermediate values: $0 < \beta < 2$.  If dust grains grow large enough to put most of the mass in bodies larger than pebbles, it should have an observable signature at these wavelengths.  Because the radiation does not penetrate far below the surface of a large body, meaning most of the material never interacts with it, the absolute value of the opacity per unit mass must decrease as the particles grow.

Calculations of $\beta$ and $\kappa_\nu$ for spherical particles show that they decrease as the particles grow (Miyake and Nakagawa 1993; Kr\"ugel \& Siebenmorgen 1994; Henning et al. 1995).  Figure 3 illustrates the mass opacity coefficients, $\kappa_\nu(a)$, for spheres of radii distributed as a power law with different maximum sizes, $a_{max}$.  In addition to the resonant absorption by silicates at 10 and 20$\mu$m, ice at 3$\mu$m, and possibly other solid-state features, the opacity is a power-law with $\beta=2$ for $\lambda > 100\,\mu$m.  For $\lambda >> a$ (Rayleigh scattering), $\kappa_\nu(a)$ is independent of size $a$, whereas for $\lambda << a$ (i.e., in the geometrical cross-section regime), $\kappa_\nu(a)$ is independent of $\nu$ but has $1/a$-dependence.  For $\lambda \simeq a$, $\kappa_\nu(a)$ is enhanced in comparison with an extrapolation from the Rayleigh-scattering regime, and $\kappa_\nu(a)$ deviates from a power-law.  Including the contribution of large particles with $a \simeq 1$\,mm or larger, Miyake and Nakagawa (1993) shows that $\beta \leq 1$ at millimeter wavelengths.  The absolute value of $\kappa_\nu$ can also be enhanced by particles with sizes similar to the observing wavelength. 

\vskip 0.1in
{
\centerline{\epsfxsize=4.25in \epsfbox{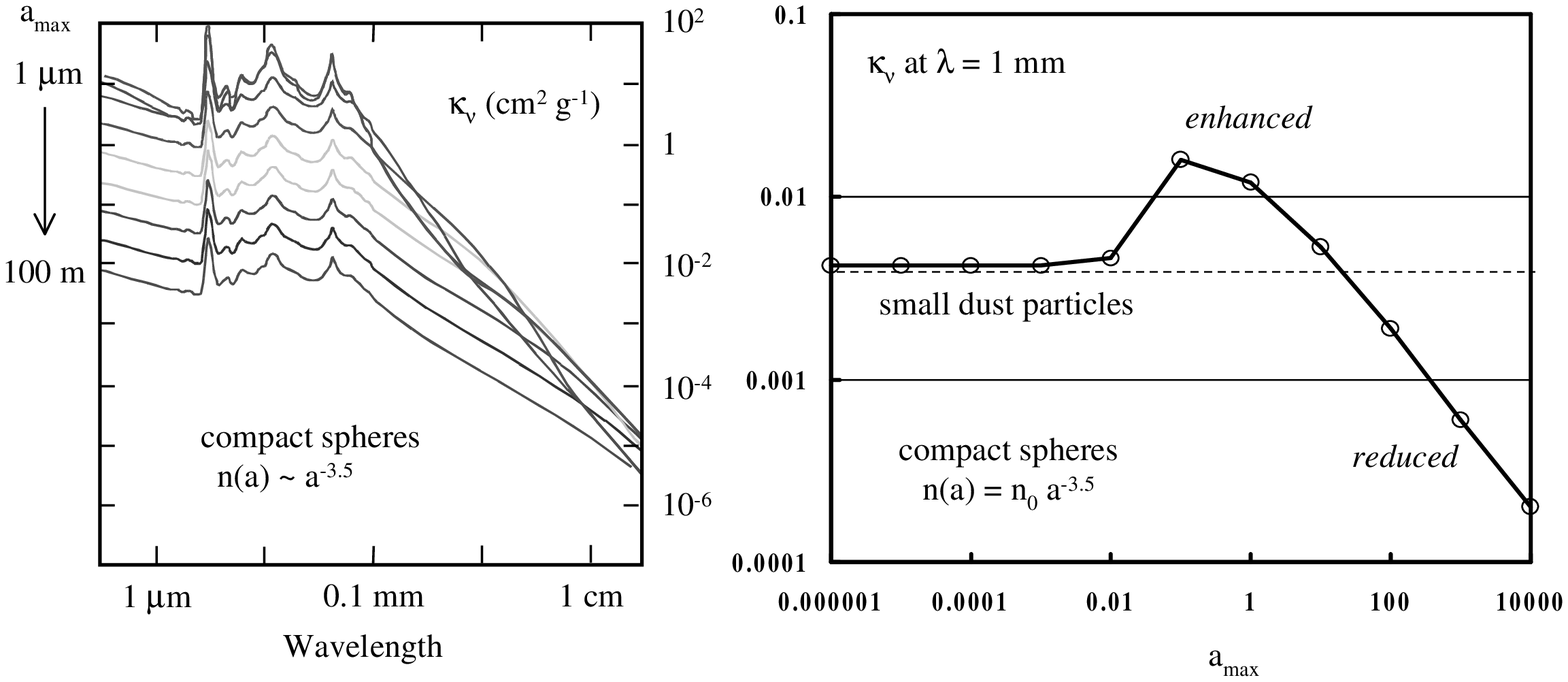}}
\vskip 0.1in
\caption{Figure 3.\capskip The left hand side shows how the mass opacity coefficient, $\kappa_\nu$, varies as a function of the maximum radius, $a_{max}$, for a collection of spherical particles, whose number density, $n$, is a power law in particle radius, $a$, with a maximum radius, $a_{max}$.  The changing slopes at long wavelengths is equivalent to a change in $\beta$.  The right hand side shows the absolute value of $\kappa_\nu$ at a wavelength of 1\,mm as the maximum particle radius changes from dust (0.01\,$\mu$m) to boulders (100\,m).  After Figures 5 \& 6 of Miyake \& Nakagawa (1993).}
}
\vskip 0.15 in

Pollack et al. (1985, 1994) proposed a model of dust particles which includes troilite (FeS) and some organics as major species in addition to silicates, water-ice, and metallic iron.  They found variations in the behaviour of $\kappa_\nu$ in the millimeter range, depending on the contributions of different constituents; the opacity is generally dominated by silicates but troilite and organic components can be more important than the silicates over restricted wavelength ranges.  Opacity enhancements can also occur in specific materials as the result of lattice resonances.  Since both changes in the particle chemistry and particle size can produce observable changes in the bulk opacities, it is not easy to demonstrate particle growth simply on the basis of changes in the opacities (Ossenkopf \& Henning 1994, Henning et al. 1995).

\subsection{B.~~Laboratory measurements of particle opacity}
The interpretation of observations of circumstellar disks depends on a knowledge of the mass absorption coefficients of particle materials at the appropriate wavelengths, generally sub-millimeter to centimeter.  It is only recently that laboratory groups have measured optical data for cosmic dust analogues at the temperatures and wavelengths appropriate for circumstellar disks.  Agladze et al. (1995, 1996) measured absorption spectra of crystalline enstatite and forsterite grains and amorphous silicate grains with $a \simeq 0.1 - 1\,\mu$m at $T = 1.2 - 30$\,K for $\lambda = 0.7-2.9$\,mm and found that the mass opacity coefficients, $\kappa_\nu$, for amorphous silicates were up to about ten times the values that are usually adopted for interstellar dust particles (Draine and Lee 1984).  The power-law behaviour also varies with $T$: $\beta$ varies between 1.5 - 2.5 for the amorphous 2MgO$\cdot$SiO$_2$ and MgO$\cdot$SiO$_2$ but has a nearly constant value of 1.2 independent of temperature for amorphous MgO$\cdot$2SiO$_2$. This behaviour comes from resonant absorption by low-lying two-level lattice states whose excitation depends on the temperature; such states are overlooked in the classic theory of solids that predicts $\beta = 2$.  Thus, in addition to the other complicating factors determining mass opacity -- size, shape, clustering effects, the presence of mantles, etc. -- the material structure is enough to alter the observable properties.  This result will complicate the interpretation of observations.  Data on amorphous and crystalline silicates (J\"ager et al. 1994, Dorschner et al. 1995, Mutschke et al. 1998, J\"ager et al. 1998), oxides (Henning et al. 1995), sulfides (Begemann et al. 1994), and ices (Hudgins et al. 1993, Preibisch et al. 1993, Ehrenfreund et al. 1997) are of special importance for protoplanetary accretion disks (see Henning and Mutschke 1997 and Mennella et al. 1998 for a discussion of the temperature effects).  Electronic data bases exist for ices and refractory solids and can be accessed by http://www.strw.leidenuniv.nl (ices) and http://www.astro.uni-jena.de (refractory solids). 

\mainsection{{V}.~~Do we see particle growth in disks?}
Circumstellar disks should contain large particles and planetesimals.  There is little doubt that many observed disks are a few Myr old, and it should take only a few percent of this time to grow rocks from tiny dust grains.  By a few Myr, much of the solid matter could be tied up in rocks and larger bodies, and we have seen that a shift in the mean particle sizes will have observable consequences for the spectral energy distributions.  Do we see these effects?

Changes in the spectral energy distributions (SEDs) of disks relative to those typical of interstellar clouds {\it are} observed (Beckwith and Sargent 1991; Mannings and Emerson 1994; Koerner et al. 1995).  The flux densities, $F_\nu$, of disks tend to fall more slowly at wavelengths longer than about 400\,$\mu$m than those of interstellar clouds, with the spectral index, $\alpha$ - where $F_\nu \propto \nu^\alpha$ - typically between 2 and 3, whereas $\alpha \sim 4$ for the interstellar medium.  Assuming optically thin emission in the Rayleigh-Jeans limit, the exponent, $\beta$, is often between 0 and 1 for disks compared to the theoretical value of 2 for small particles in the long wavelength limit.  The right hand part of Figure 4 shows the distribution of emissivity exponents, $\beta$, determined from the observed spectral slope of disk emission near $\lambda ~ 1$\,mm.  Many of the disks in this sample have $\beta \le 1$.  It is tempting to interpret this change as the result of particle growth.  

Unfortunately, there are many adjustable parameters in the disk models, and  several different parameter combinations usually produce acceptable fits to the same SEDs (Thamm, Steinacker, and Henning 1994).  A model typically has the following parameters: $r_0$, $R_D$, $\kappa_\nu$, $\theta$ (inclination to line of sight), $T_0$ and $q$ ($T(r) = T_0 (r / r_0)^{-q}$), $\Sigma_0$ and $p$ ($\Sigma(r) = \Sigma_0 (r / r_0)^{-p}$).  There are rarely more than ten data points in a SED to be fitted.  The SEDs are degenerate with respect to the models and almost certainly inadequate by themselves to demonstrate grain growth in disks.

Fortunately, some of these parameters affect the millimeter-wave\-length emission weakly or can be determined reasonably well from other wavelengths.  The inner radius of the disk, $r_0$, is unimportant for wavelengths beyond 20\,$\mu$m in almost all cases.  Both $T_0$ and $q$ are well constrained by the optically thick emission between 10 and 100\,$\mu$m.  The disk inclination, $\theta$, is unimportant for the optically thin (mm-$\lambda$) emission but important for the optically thick ($\lambda < 100\,\mu$m) part of the SED.  In some cases, $\theta$ can be estimated directly from images.  

The remaining parameters -- $\Sigma_0$, $p$, $R_D$, $\kappa_\nu$ -- play a role in the long wavelength emission and, hence, the derivation of the emissivity exponent, $\beta$, from the observed spectral slope, $\alpha$.  In the simplest case of $p = 0$ (constant surface density), $\tau_\nu < 1$ for all wavelengths longer than 100\,$\mu$m, say, and $T(r)$ large enough for all $r$ to make the Rayleigh-Jeans approximation valid for millimeter wavelengths, $\beta = \alpha - 2$ (see Eqn. 10), the spectral index gives the emissivity exponent directly.  But if the density decreases at least as fast as $r^{-{3\over 2}}$ ($p = {3\over 2}$), there is substantial contribution to the long wavelength emission from the inner parts of the disk.  If $p > {3\over 2}$, the inner parts of the disk dominate the emission making $R_D$ irrelevant but implying that most of the emission at any wavelength comes from radii near the radius at which the disk becomes optically thick (see Beckwith et al. 1990 for a detailed discussion of these effects).  In this case, the spectral index, $\alpha$, depends on $\beta$ {\it and} both $p$ and $q$.  Specifically, Beckwith et al. (1990) derived $\beta$ in terms of the model parameters as:

$$ {\beta \approx (\alpha - 2) (1 + {p \over (2 - q) \ln \left[ {2 \over (2 - p) \bar\tau_\nu} \right]} ) }, \eqno(11) $$

\noindent where $\bar\tau_\nu = {\kappa_\nu M_D \over \cos\theta \pi R_D^2}$ is the average optical depth at frequency, $\nu$, and $M_D$ is the mass of the disk.\footnote{\ddag}{Beckwith et al. derive the spectral index from $\nu F_\nu$, not $F_\nu$ used here.}  Equation 11 depends strongly on $p$ which cannot be determined from SEDs alone.   For this reason, the emissivity exponent derived from and SED is model dependent. 

The right hand part of Figure 4 includes the distribution of $\beta$'s derived from model fits assuming the surface density is proportional to $r^{-3/2}$.  Including the surface density distribution tends to increase the derived value of the emissivity exponent relative to the observed spectral slope.  For example for 1.3\,mm observation of a disk with $M_D = 0.03\,$M$_\odot$, $R_D = 100\,$AU, and $\kappa_\nu = 0.02\,$cm$^2$\,g$^{-1}$, the average optical depth, $\bar\tau_\nu = 0.24$, and $\beta \approx 1.36(\alpha - 2)$.  Any conclusions about particle growth depend on knowledge of how much optically thick parts of the disks contribute to the spectral energy distribution at long wavelengths.  It is probably for this reason that the samples of disks with measured $\alpha$'s at long wavelengths have not be enlarged much since the work of Beckwith \& Sargent (1991) and Mannings \& Emerson (1994).

To measure the optical depth, it is essential that the disk emission be resolved at wavelengths where it is relatively transparent so that the surface density distribution may be determined directly from the distribution of optical depth.  Resolving the disks is time-consuming with the present suite of millimeter facilities.  Several groups have now managed to resolve the emission from one star, HL~Tau, by carrying out novel experiments with existing millimeter-wave telescopes.  Lay et al. (1994) and Lay, Carlstrom \& Hills (1997) combined the CSO and JCMT into a single-baseline interferometer to observe at 0.65 and 0.87\,mm and combined it with data from the Owens Valley millimeter interferometer at 1.3\,mm.  Mundy et al. (1996) extended the BIMA array to measure a size of the disk at 2.7\,mm wavelength, and  Wilner, Ho \& Rodriguez (1996) used the VLA to get sub-arcsecond resolution of the disk at 7\,mm.  Each group resolved the emission but with beam sizes of the same order as the disk size, and the sub-millimeter observations had $u-v$ plane coverage too poor to map the brightness distribution.  All groups measured the average brightness temperature and overall disk orientation and constrained subsequent model fitting much more than is possible from the SEDs alone.

\vskip 0.1in
{
\centerline{\epsfxsize=4.25in \epsfbox{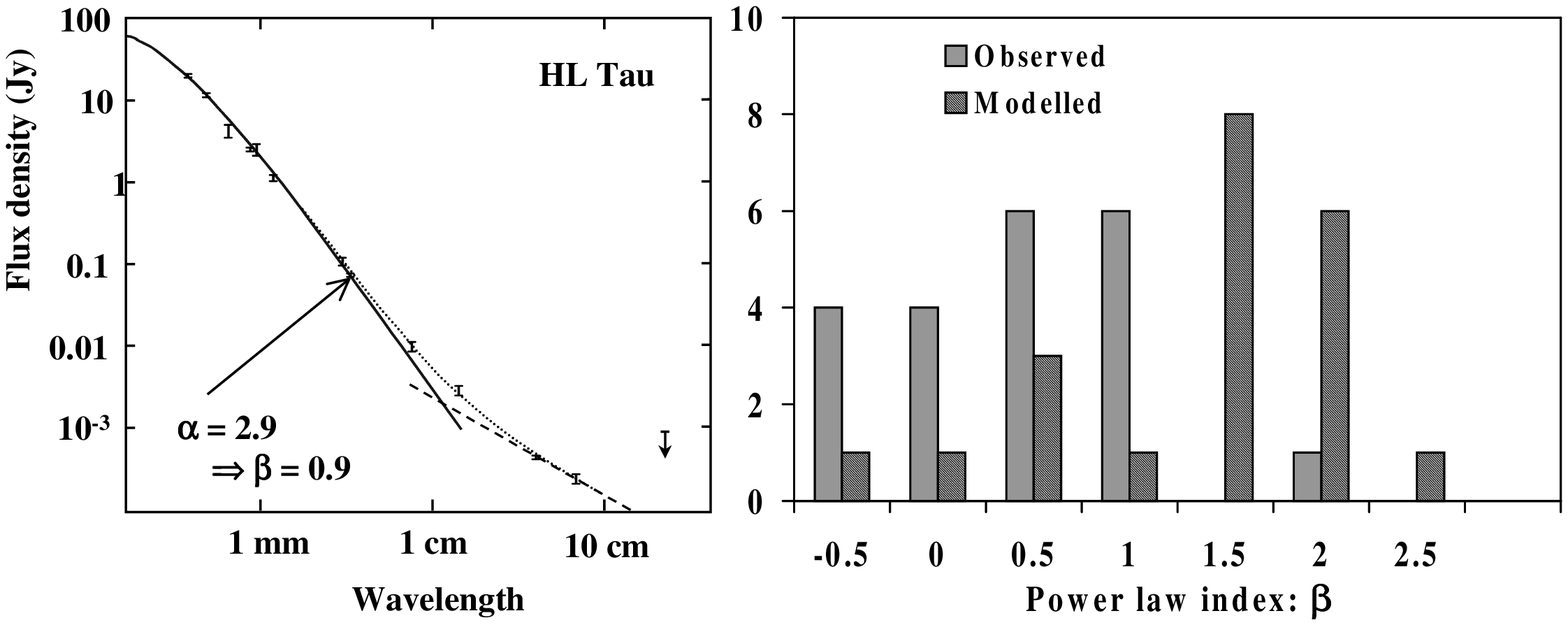}}
\vskip 0.1in
\caption{Figure 4.\capskip The left hand part shows the spectral energy distribution of HL Tau from about 350\,$\mu$m to 6\,cm wavelength. The solid line fits the short wavelengths -- thermal emission by dust -- and and the dashed line fits the long wavelengths -- free-free emission from ionized gas.  The long wavelength spectral index of the dust is 2.9, meaning $\beta = 0.9$ with no corrections for optically thick emission (Wilner, Ho, \& Rodriguez 1996).  The right hand portion shows the distribution of $\beta$'s derived in in two ways for a larger sample of disks.  The ''observed'' $\beta$'s are derived directly from the spectral index as described in the text: $\beta_{obs} = \alpha - 2$.  The ''model'' $\beta$'s are derived by assuming a surface density distribution as discussed in the text (Beckwith \& Sargent 1991).}
}
\vskip 0.15 in

Unfortunately, the combined data are not consistent with simple power-law models for disk optical depth.  The apparent size of the HL~Tau disk is approximately constant from 0.65\,mm to 7\,mm.  Because the optical depths should drop by factors of 10 (for $\beta = 1$) to 100 ($\beta = 2$) over this wavelength range, any centrally concentrated density distribution, $p \ge 1$, would immediately imply smaller sizes at longer wavelengths.  Lay, Carlstrom \& Hills (1997) discuss the modifications in the models that might reconcile the data.  They conclude that a mixture of particle sizes such that there is no unique value of $\beta$ may explain the data.  We note that there would have to be substantial contributions to the emission from particles of order 1\,cm for this conclusion to hold.  Although such particles are not yet the size of planetesimals, they are more than four orders of magnitude larger than typical interstellar grains and would be present only if there was substantial particle growth in the disk.  HL~Tau is one of the youngest disks (Cohen 1983; Beckwith et al. 1990), so demonstration of particle growth in this system would imply that it occurs quickly as theory suggests.

HL~Tau is also one of the more complicated star/disk systems to interpret, owing to the large amount of surrounding material and the likelihood that high optical depths are contributing to the millimeter-wave emission (cf. Stapelfeldt et al. 1995; Beckwith and Birk 1995; Weintraub et al. 1995; Cabrit et al. 1996; Menshchikov et al. 1998).  Mundy et al. (1996) observe a brightness temperature of 29\,K from the HL~Tau disk at 2.7\,mm with an outer radius of order 150\,AU.  This temperature is very close to the physical temperature of the outer regions of the disk: Beckwith et al. (1990) derive $T = 32$\,K at a disk radius of 100\,AU.  Since the brightness temperature, $T_B$, is related to the physical temperature by $T_B = T (1 - e^{-\tau})$, there must be a substantial contribution from optically thick dust that is not easily taken into account in the model fits.

Because HL~Tau is much younger, more massive, and considerably more complex than most of the disks in Figure 4, the derived $\beta$'s may still be valid in many cases.  But since the first direct measurement of the optical depth distribution show the model assumptions are invalid, we cannot conclude yet that the long wavelength spectral indices tell us much about the particle properties. 

The evidence for grain growth is, therefore, inconclusive.  There are quite a few disks with spectral indices that are inconsistent with optically thin emission from very small grains.  The spectral indices are not unambiguous indicators of large grains, however. Contributions from high optical depth emission within the disks can flatten the SEDs in similar ways to changes in the grain-size distribution.  Furthermore, $\beta = 1$ is {\it not} sufficient to demonstrate grain growth.  Certain kinds of materials can produce an index of 1 even for small grains.  There are quite a few disks for which the spectral index implies $\beta < 1$, but none have been resolved at millimeter-wavelengths.  

These observations are difficult, because the required spatial resolution is much higher than that routinely available with existing milli\-meter-wave interferometers.  This situation will change during the next decade.  The proposed construction of a large millimeter-wave interferometer in the southern hemisphere - the American MMA and European LSA projects - will yield a combination of sensitivity, resolution, and wavelength coverage allowing all the disks in the dark clouds within $\le 300$\,pc to be easily resolved.  Mapping the disk emission should provide a direct measure of the optical depth, assuming the physical temperature can be observed independently, from molecular lines, say.  From the optical depth at different wavelengths, it should be possible to demonstrate that a population of large grains or rocks does contribute to the emission, if such a population exists.  This avenue is the most obvious hope for a demonstration of grain growth in disks.

In the context of planet formation, it is gratifying that the evidence favors grain growth in some disks, even if this evidence is insufficient to establish unambiguously the presence of large particles.  The situation is similar to that of the disks themselves in the last decade.  At that time, many observations were consistent with the existence of circumstellar disks with the right properties for planet formation, but the evidence was not insufficient for proof beyond reasonable doubt.  Now, there is no doubt that disks exist with the general properties then inferred, and a separate line of research has uncovered planets themselves - albeit indirectly - orbiting other stars.  It is natural to suppose that disks are the necessary precursors to planets.  The next generation of millimeter-wave observatories may well prove the evolutionary link by demonstrating particle growth, too.


\vskip .25in
\centerline{\bf REFERENCES}

\ref{Adams, F.C., Emerson, J.P., and Fuller, G.A. 1990. Submillimeter
photometry and disk masses of T Tauri disk systems. {\refit Astrophys. J.\/}
357:606--620.}

\ref{Adams, F.C., and Shu, F.H. 1986. Infrared spectra of rotating protostars. {\refit Astrophys. J.\/} 308:856--833.}

\ref{Adams, F.C., Lada, C.J., and Shu, F.H. 1988. The disks of T Tauri stars with flat infrared spectra. {\refit Astrophys. J.\/} 326:865--883.}

\ref{Adams, F.C., and Watkins, R. 1995. Vortices in circumstellar disks. {\refit Astrophys. J.\/} 451:314-327.}

\ref{Agladze, N. I., Sievers, A. J., Jones, S. A., Burlitch, J. M., and Beckwith, S. V. W. 1995. Reassessment of millimetre-wave absorption coefficients in interstellar silicate grains. {\refit Nature\/}, 372:243-245.}

\ref{Agladze, N. I., Sievers, A. J., Jones, S. A., Burlitch, J. M., \& Beckwith, S. V. W. 1996. Laboratory results on millimeter-wave absorption in silicate grain materials at cryogenic temperatures. {\refit Astrophys. J.\/}, 462:1026--1040.}

\ref{Andr\'e, P., and  Montmerle, T. 1994. From T Tauri stars to protostars: Circumstellar material and young stellar objects in the rho Ophiuchi cloud. {\refit Astrophys. J.\/}, 420:837-862.}

\ref{Appenzeller, I., Jankovics, I., and \" Ostreicher, R. 1984. Forbidden-line profiles of T Tauri stars. {\refit Astron. Astrophys.\/} 141:108--115.}

\ref{Balbus, S.A., and Hawley, J.F. 1991a. A powerful local shear instability in weakly magnetized disks. I. Linear analysis. {\refit Astrophys. J.\/} 376:214--222.}

\ref{Balbus, S.A., and Hawley, J.F. 1991b. A powerful local shear instability in weakly magnetized disks. II. Nonlinear evolution. {\refit Astrophys. J.\/} 376:223--233.}

\ref{Barge, P., and Sommeria, J. 1995. Did planet formation begin inside persistent gaseous vortices? {\refit Astron. Astrophys.\/} 295:L1-L4.}

\ref{Beckwith, S. V. W., and Birk, C. 1995. Vertical disk structure in HL Tauri. {\refit Astrophys. J.\/}, 449:L59--L63.}

\ref{Beckwith, S. V. W. and Sargent, A. I. 1991. Particle emissivity in circumstellar disks. {\it Astrophys. J.\/}, 381:250-258.}

\ref{Beckwith, S.V.W., Sargent, A.I., Chini, R.S., and G\"usten, R. 1990. A survey for circumstellar disks around young stellar objects. {\refit Astron. J.\/} 99:924--945.}

\ref{Begemann, B., Dorschner, J., Henning, Th., Mutschke, H. and Thamm, E. 1994. A laboratory approach to the interstellar 
sulfide dust problem. {\ref Astrophys. 
J.\/} 423:L71-L74.}

\ref{Bell, K.R., Cassen, P.M., Klahr, H.H., and Henning, Th. 1997. The structure and appearance of protostellar accretion disks: limits on disk flaring. {\refit Astrophys. J.\/} 486:372-387.}

\ref{Blum, J., Wurm, G., Poppe, T., and Heim, L.-O. 1998. Aspects of laboratory dust aggregation with relevance to the formation of planetesimals.  In {\refit Laboratory Astrophysics and Space Research\/}, eds.\  P. Ehrenfreund, and H. Kochan (Dordrecht: Kluwer Acad. Publ.), in press.}

\ref{Blum, J., Wurm, G., Kempf, S., and Henning, Th. 1996. The Brownian motion of dust  in the solar nebula: an experimental approach to the problem of pre-planetary dust aggregation. {\refit Icarus\/} 124:441-451.}

\ref{Bodenheimer, P. 1995, {\refit Ann. Rev. Astron. Astroph.}, 33:199-238.}

\ref{Bohren, C. F., and Huffman, D. R. 1983. {\refit Absorption and Scattering of Light by Small Particles} (New York:Wiley).}

\ref{Brown, W., and Ball, R. 1985. Computer simulation of chemically limited aggregation. {\refit J. Phys. A\/} 18:L517-L521.}

\ref{Brandenburg A., and  Hodgson, L.S. 1998. Turbulence effects in planetesimal formation. {\refit Astron. Astrophys.\/} 330:1169-1174.} 

\ref{Bridges, F.G., Supulver, K.D., Lin, D.N.C., Knight, R., and Zafra, M. 1996. Energy loss and sticking mechanisms in particle aggregation in planetesimal formation. {\refit Icarus\/} 123:422-435.}

\ref{Cabot, W., Canuto, V.M., Hubickyj, O., and Pollack, J.B. 1987. The role of turbulent convection in the primitive solar nebula. II. Results. {\refit Icarus\/} 69:387--422.}

\ref{Cabrit, S., Guilloteau, S., Andr\'e, P., Bertout, C., Montmerle, T., and Schuster, K. 1996. Plateau de Bure observations of HL Tauri: outflow motions in a remnant circumstellar envelope. {\refit Astron Astrophys.\/}, 305:527-540.}

\ref{Calvet, N., Hartmann, L., Kenyon, S. J., and Whitney, B. A. 1994, {\refit Astrophys. J.\/}, 434:330-340.}

\ref{Cameron, A. G. W. 1988. Origin of the solar system. {\it Ann. Rev. Astron. Astrophys.}, 26:441-472.}

\ref{Cameron, A.G.W., Fegley, M.B. 1982. Nucleation and condensation in the primitive solar nebula. {\it Icarus\/}, 52:1-13.}

\ref{Champney, J.M., Dobrovolskis, A.R., Cuzzi, J.N. 1995. A numerical turbulence model for multiphase flows in the protoplanetary nebula. {\refit Phys. Fluids\/} 7:1703-1711.}

\ref{Chiang, E., and Goldreich, P. 1997. Spectral energy distributions of T Tauri stars with passive circumstellar disks. {\refit Astrophys. J.\/}, 490:368-376.}

\ref{Chokshi, A., Tielens, A.G.G.M., and Hollenbach, D. 1993. Dust coagulation. {\refit Astrophys. J.\/} 407:806-819.}

\ref{Cohen, M. 1983, {\refit Astrophys. J. \/} 270:L69--L71.}

\ref{Cuzzi, J.N., Dobrovolskis, A.R., and Champney, J.M. 1993. Particle-gas dynamics in the midplane of a protoplanetary nebula. {\refit Icarus\/} 106:\-102-134.}

\ref{Cuzzi, J.N., Dobrovolskis, A.R., and Hogan, R.C. 1996.  Turbulence, chondrules, and planetesimals. In {\refit Chondrules and the Protoplanetary Disk\/}, eds.\ R. Hewins, R. Jones, and E.R.D. Scott (Cambridge: Cambridge Univ. Press), pp.\ 35--43.}

\ref{Dominik, C., and  Tielens, A.G.G.M. 1997. The physics of dust coagulation and the structure of dust aggregates in space. {\refit Astrophys. J.\/} 480:647-673.}

\ref{Dorschner, J., Begemann, B., Henning, Th., J\"ager, C., and Mutschke, H. 1995. Steps towards interstellar silicate mineralogy. II. Study of Mg-Fe silicate glasses of variable composition. {\refit Astron. Astrophys.\/} 300:503-520.}


\ref{Draine, B. T., and Lee, H. M. 1984. Optical properties of interstellar graphite and silicate grains. {\refit Astrophys. J.}, 285:89-108.}

\ref{Dubrulle, B. 1992. A turbulent closure model for thin accretion disks. {\refit Astron. Astrophys.\/} 266:592--604.}

\ref{Dubrulle, B. 1993. Differential rotation as a source of angular momentum transfer in the solar nebula. {\refit Icarus\/} 106:59--76.}

\ref{Dubrulle, B., Morfill, G., and  Sterzik, M. 1995. The dust subdisk in the protoplanetary nebula. {\refit Icarus\/} 114:237-246.}

\ref{Edwards, S., Cabrit, S., Strom, S., Heyer, I., Strom, K., and Anderson, E. 1987. Forbidden line and H$\alpha$ profiles in T Tauri spectra: a probe of anisotropic mass outflows and circumstellar disks. {\refit Astrophys. J.\/} 321:473--495.}

\ref{Ehrenfreund, P., Boogert, A.C.A., Gerakines, P.A., Tielens, A.G.G.M., and  van Dishoeck, E.F. 1997. Infrared spectrsocopy of interstellar apolar ice analogs. {\refit Astron. Astrophys.\/} 328:649-671.}

\ref{Emerson, J. P. 1988. In {\refit  Formation and Evolution of Low Mass Stars\/}, eds. A. K. Dupree, and  M. T. V. Lago (Dordrecht: Kluwer Acad. Publ.), pp.21-44.}

\ref{Finocchi, F., Gail, H.-P., and Duschl, W. 1997. Chemical reactions in protoplanetary disks. II. Carbon dust oxidation. {\refit Astron. Astrophys.\/} 325:1264-1279.}

\ref{Gail, H.-P. 1998. Chemical reactions in protoplanetary accretion disks. IV. Multicomponent dust mixture. {\refit Astron. Astrophys.\/} 332:1099-1122.}


\ref{Goldreich, P., and Ward, W.R. 1973. The formation of planetesimals. {\refit Astrophys. J.\/} 183:1051-1061.}

\ref{Grossman, L. 1972. Condensation in the primitive solar nebula. 
{\refit Geo\-chim. Cosmochim. Acta\/} 38:47-64.}

\ref{Hayashi, C. 1981. Structure of the solar nebula, growth and decay of magnetic fields and effects of magnetic and turbulent viscosities on the nebula. {\refit Prog. Theor. Phys. Suppl.} 70:35--53.}


\ref{Henning, Th., Begemann, B., Mutschke, H., and  Dorschner, J. 1995. Optical properties of oxide dust grains. {\refit Astron. Astrophys. Suppl. Ser.\/} 112:143-149.

\ref{Henning, Th., Michel, B., and Stognienko, R. 1995. Dust opacities in dense regions. {\refit Planetary \& Space Science}, 43:1333-1343.}

\ref{Henning, Th., and  Mutschke, H. 1997. Low-temperature infrared properties of cosmic dust analogues. {\refit Astron. Astrophys.\/} 327:743-754.}

\ref{Henning, Th., Pfau, W., Zinnecker, H., and  Prusti, T. 1993. A 1.3 mm survey for circumstellar dust around young Chamaeleon objects.  {\refit Astron. Astrophys.}, 276:126-138.}

\ref{Henning, Th., and  Stognienko, R. 1996. Dust opacities for protoplanetary accretion disks - influence of dust aggregates. {\refit Astron. Astrophys.\/} 311:291-303.}

\ref{Hewins, R.H., Jones, R.H., and  Scott, E.R.D., eds., 1996. {\refit Chondrules and the Protoplanetary Disk\/} (Cambridge: Cambridge Univ. Press)}

\ref{Hildebrand, R. H. 1983. The determination of cloud masses and dust characteristics from submillimetre thermal emission. {\refit Q.J. RAS\/}, 24:267.}


\ref{Hudgins, D. M., Sandford, S. A., Allamandola, L. J., and  Tielens, A. G. G. M. 1993. Mid- and far-infrared spectroscopy of ices: optical constants and integrated absorbances. {\refit Astrophys. J. Suppl. Ser.} 86:713-870.}

\ref{Hutchins, D.K., Harper, M.H., and  Felder, R.L. 1995. Slip correction measurements for solid spherical particles by modulated dynamic light scattering. {\refit Aerosol. Sci. Technol.\/} 22:202-218.}

\ref{J\"ager, C., Molster, F.J., Dorschner, J., Henning, Th., Mutschke, H., and  Waters, L.B.F.M.  1998. Steps towards interstellar silicate mineralogy. IV. The crystalline revolution {\refit Astron. Astrophys.\/} 339:904-916.}

\ref{J\"ager, C., Mutschke, H., Begemann, B., Dorschner, J., and  Henning, Th. 1994. Steps towards interstellar silicate mineralogy. I. Laboratory results of a silicate glass of mean cosmic composition. {\refit Astron. Astrophys.\/} 292:641-655.}

\ref{Kempf, S., Pfalzner, S., and  Henning, Th. 1998. N-particle simulations of dust growth: I. Growth driven by Brownian motion. {\refit Icarus}, submitted.}

\ref{Kenyon, S. J., and Hartmann, L. 1987, {\refit Astrophys. J. \/}323:714--733.}

\ref{Klahr, H.H., and Henning, Th. 1997. Particle-trapping eddies in protoplanetary accretion disks. {\refit Icarus\/} 128:213-229.}

\ref{Klahr, H. H., Henning, Th., and  Kley, W. 1998. On the azimuthal structure of thermal convection in circumstellar disks. {\refit Astrophys. J.}, in press.} 

\ref{Kley, W., Papaloizou, J.C.B., and Lin, D.N.C. 1993. On the momentum transport associated with convective eddies in accretion disks. {\refit Astrophys. J.\/} 416:679--689.}

\ref{Koerner, D. W., Chandler, C. J., and Sargent, A. I. 1995. Aperture synthesis imaging of the circumstellar disk around DO Tauri. {\refit Astrophys. J.}, 452:L69-L72.}

\ref{Kr\"ugel, E., and Siebenmorgen, R. 1994. Dust in protostellar cores and stellar disks. {\refit Astron. Astrophys.}, 288:929.}


\ref{Lay, O. P., Carlstrom, J. E., Hills, R. E., and Phillips, T. G. 1994, {\refit Astrophys. J.\/}, 434:L75.}

\ref{Lay, O. P., Carlstrom, J. E., and Hills, R. E. 1997, {\refit Astrophys. J.\/}, 489:917-927.}

\ref{Lissauer, J. 1993. Planet formation. {\refit Ann. Rev. Astron. Astrophys.}, 31:129-174}

\ref{Lunine, J. I., Engel, S., Rizk, B., and  Horanyi, M. 1991. Sublimation and reformation of icy grains in the primitive solar nebula. {\refit Icarus\/} 94:333-344.}

\ref{Lynden-Bell, D., and Pringle, J. E. 1974. The evolution of viscous discs and the origin of the nebular variables. {\refit MNRAS\/}, 168:603-637.}

\ref{Mannings, V., and Emerson, J.P. 1994. Dust in discs around T Tauri stars: 
grain growth? {\refit Mon. Not. Roy. Astron. Soc.\/} 267:361--378.}

\ref{McCaughrean, M. J., and O'Dell, C. R. 1996. Direct Imaging of Circumstellar Disks in the Orion Nebula. {\refit Astron. J.\/}, 111:1977-1986.}

\ref{Meakin, P. 1984: Effects of cluster trajectories on cluster-cluster aggregation: a comparison of linear and Brownian trajectories in two- and three-dimensional simulations. {\refit Phys. Rev. A\/} 29:997-999.}

\ref{Meakin, P., Donn, B., and  Mulholland, G. 1989. Collisions between point masses and fractal aggregates. {\refit Langmuir\/} 5:510-518.}

\ref{Mennella, V., Brucato, J. R., Colangeli, L., Palumbo, P., Rotundi, A., and  Bussoletti, E. 1998. Temperature Dependence of the Absorption Coefficient of Cosmic Analog Grains in the Wavelength Range 20 Microns to 2 Millimeters. {\refit Astrophys. J.}, 496:1058-1066.}

\ref{Menshchikov, A. B., Henning, Th., and Fischer, O. 1998. Self-con\-sis\-tent model of the dusty torus around HL~Tau. {\refit Ap. J.}, submitted.}

\ref{Miyake, K., and Nakagawa, Y. K. 1993. Effects of particle size distribution on opacity curves of protoplanetary disks around T Tauri stars. {\refit Icarus\/} 106:20--41.}


\ref{Mizuno, H. 1989. Grain growth in the turbulent accretion disk solar nebula. {\refit Icarus\/} 80:189-201.}

\ref{Mizuno, H., Markiewicz, W.J., and V\"olk, H.J. 1988. Grain growth in turbulent protoplanetary accretion disks. {\refit Astron. Astrophys.\/} 195:183-192.}

\ref{Morfill, G.E. 1988. Protoplanetary accretion disks with coagulation and evaporation. {\refit Icarus\/} 75:371-379.} 

\ref{Mundy, L. G. et al. 1996. Imaging the HL Tauri Disk at lambda = 2.7 Millimeters with the BIMA Array. {\refit Astrophys. J.\/}, 464:L169-L173.}

\ref{Mutschke, H., Begemann, B., Dorschner, J., G\"urtler, J., Gustafson, B., Henning, Th., and Stognienko, R. 1998. Steps towards interstellar silicate mineralogy. III. The role of aluminium in circumstellar amorphous silicates. {\refit Astron. Astrophys.\/} 333:188-198.

\ref{Nakagawa, Y.K., Nakazawa, K., and  Hayashi, C. 1981. Growth and sedimentation of dust grains in the primordial solar nebula. {\refit Icarus\/} 45:517-528.}

\ref{Nakagawa, Y., Sekiya, M., and  Hayashi, C. 1986. Settling and growth of dust particles in a laminar phase of a low-mass solar nebula. {\refit Icarus\/} 67:355--390.

\ref{Nakamoto, T., and Nakagawa, Y. 1994. Formation, early evolution, and gravitational stability of protoplanetary disks. {\refit Astrophys. J.\/} 421:640--651.}

\ref{Nakamoto, T., and Nakagawa, Y. 1995. Growth of protoplanetary disks around young stellar objects. {\refit Astrophys. J.\/} 445:330--336.}

\ref{Neufeld, D.A.,and  Hollenbach, D.J. 1994. Dense molecular shocks and accretion onto protostellar disks. {\refit Astrophys. J.\/} 428:170-185.}

\ref{Nuth III, J.A., and Wilkinson, G.M. 1995. Magnetically enhanced coagulation of very small iron grains: A correction of the enhancement factor due to dipole-dipole interactions. {\refit Icarus\/} 117:431-434.}

\ref{Nuth III, J.A., Berg, O., Faris, J., and  Wasilewski, P. 1994. Magnetically enhanced coagulation of very small iron grains. {\refit Icarus\/} 107:155-163.}

\ref{O'Dell, C. R., and Wen, Z. 1994. Postrefurbishment mission Hubble Space Telescope images of the core of the Orion Nebula: Proplyds, Herbig-Haro objects, and measurements of a circumstellar disk.{\refit Astrophys. J.\/}, 436:194-202.}

\ref{Ossenkopf, V. 1993. Dust coagulation in dense molecular clouds: the formation of fluffy aggregates. {\refit Astron. Astrophys.\/} 280:617-646.}

\ref{Ossenkopf, V., and  Henning, Th. 1994. Dust opacities for protostellar cores. {\refit Astron. Astrophys.\/} 291:943-959.}

\ref{Papaloizou, J. C. B. and Lin, D. N. C. 1995, {\refit Ann. Rev. Astron. Astroph.}, 33:505-540.}

\ref{Pollack, J. B., Hollenbach, D., Beckwith, S., Simonelli, D.P., Roush, T., and Fong, W. 1994. Composition and radiative properties of grains in molecular clouds and accretion disks. {\refit Astrophys. J.\/} 421:615-639.}

\ref{Poppe, T., Blum, J., and Henning, Th. 1997. Generating a jet of de-agglomerated small particles in vacuum. {\refit Rev. Sci. Instrum.\/} 68:2529-2533.}

\ref{Poppe, T., Blum, J., and Henning, Th. 1998. Analogous experiments on the stickiness of micron-sized preplanetary dust. {\refit Astrophys. J..\/} submitted}

\ref{Preibisch, T., Ossenkopf, V., Yorke, H.W., and  Henning, Th. 1993. The influence of ice-coated grains on protostellar spectra. {\refit Astron. Astrophys.\/} 279:577-588.}

\ref{Pringle, J. E. 1981. Accretion discs in astrophysics. {\refit Ann. Rev. Astron. Astrophys.}, 19:137-162.}

\ref{Prinn, R.G.  1993. Chemistry and evolution of gaseous circumstellar disks. In {\refit Protostars \& Planets {I}{I}{I}\/}, eds.\ E. H. Levy and J. I. Lunine (Tucson: Univ.\ of Arizona
Press). pp.\ 1005-1028.}

\ref{Ruden, S.P., Pollack, J.B. 1991. The dynamical evolution of the protosolar nebula. {\refit Astrophys. J.\/} 375:740-760}

\ref{Safronov, V. 1969, in Evolution of the Protoplanetary Cloud and the Formation of the Earth \& Planets, NASA TTF-667, (translation from Russian).}

\ref{Safronov, V. S. 1972. {\refit Evolution of the protoplanetary cloud and formation of the earth and planets (Moscow:Nauka Press)}

\ref{Schmitt, W., Henning, Th., and Mucha, R. 1997. Dust evolution in protoplanetary accretion disks. {\refit Astron. Astrophys.\/} 325:569-584.}

\ref{Sekiya, M. 1998. Quasi-equilibrium density distributions of small dust aggregations in the solar nebula. {\refit Icarus\/} 133:298-309.}

\ref{Squires, K. D., and Eaton, J. K.. 1991. Preferential concentration of particles by turbulence. {\refit Phys. Fluids A\/} 3:1169-1178.}

\ref{Stapelfeldt, K. R. et al. 1995. WFPC2 Imaging of the Circumstellar Nebulosity of HL Tauri. {\refit Astrophys. J.}, 449:888-893.}

\ref{Stepinski, T.F., Valageas, P. 1996. Global evolution of solid matter in turbulent protoplanetary disks. I. Aerodynamics of solid particles. {\refit Astron. Astrophys.\/} 309:301-312.}

\ref{Stepinski, T.F., Valageas, P. 1997. Global evolution of solid matter in turbulent protoplanetary disks. II. Development of icy planetesimals. {\refit Astron. Astrophys.\/} 319:1007-1019.}


\ref{Stone, J.M., and Balbus, S.A. 1996. Angular momentum transport in accretion disks via convection. {\refit Astrophys. J.\/} 464:364--372.}

\ref{Strom, S. E., Edwards, W., and Skrutskie, M. F. 1993. Evolutionary time scales for circumstellar disks associated with intermediate- and solar-type stars. In {\refit Protostars \& Planets III\/}, eds. E. H. Levy, and J. Lunine (Tucson: Univ. of Arizona Press}, pp.837-866.

\ref{Supulver, K.D., Bridges, F.G., Tiscareno, S., Lievore, J., and Lin, D.N.C. 1997. The sticking properties of water frost produced under various ambient conditions. {\refit Icarus\/} 129:539-554.}


\ref{Tanga, P., Babiano, A., Dubrulle, B., and Provenzale, A. 1996. Forming planetesimals in vortices. {\refit Icarus\/} 121:158-170}

\ref{Tanaka, H., and  Nakazawa, K. 1994. Validity of the statistical coagulation equation and runaway growth of protoplanets. {\refit Icarus\/} 107:404-412.}

\ref{Thamm, E., Steinacker, J., and Henning, Th. 1994. Ambiguities of para\-metrized dust disk models for young stellar objects. {\refit Astron. Astrophys.} 287:493-502.} 


\ref{Tscharnuter, W.M., and Boss, A.P. 1993. Formation of the protosolar nebula. In {\refit Protostars \& Planets {I}{I}{I}\/}, eds.\ E. H. Levy and J. I. Lunine (Tucson: Univ.\ of Arizona Press), pp.\ 921--938.} 

\ref{V\"olk, H.J., Jones, F.C., Morfill, G.E., and  R\"oser, S. 1980. Collisions between grains in a turbulent gas. {\refit  Astron. Astrophys.\/} 85:316-325.}



\ref{Weidenschilling, S.J. 1977. Aerodynamics of solid bodies in the solar nebula. {\refit Mon. Not. Roy. Astron. Soc. \/} 180:57-70.} 

\ref{Weidenschilling, S. J. 1980. Dust to planetesimals: Settling and coagulation in the solar nebula. {\refit Icarus\/} 44:172-189.} 

\ref{Weidenschilling, S. J. 1988. Formation processes and time scales for meteorite parent bodies, In {\refit  Meteorites and the early solar system\/}  (Tucson: Univ.\ of Arizona Press). pp.\ 348--371.}

\ref{Weidenschilling, S. J. 1995. Can gravitational instability form planetesimals? {\refit Icarus\/} 116:433-435.} 

\ref{Weidenschilling, S. J. 1997. The origin of comets in the solar nebula. {\refit Icarus\/} 127:290-306.} 

\ref{Weidenschilling, S.J., and Cuzzi, J.N. 1993. Formation of planetesimals in the solar nebula. In {\refit Protostars \& Planets {I}{I}{I}\/}, eds.\ E. H. Levy and J. I. Lunine (Tucson: Univ.\ of Arizona Press). pp.\ 1031-1060.}

\ref{Weintraub, D. A., Kastner, J. H., and  Whitney, B. A. 1995, {\refit Astrophys. J.}, 452:L141-L145.}

\ref{Wilner, D. J., Ho, P. T. P., and  Rodriguez, L. F. 1996. Subarcsecond VLA observations of HL Tauri: Imaging the circumstellar disk. {\refit Astrophys. J.}, 470:L117-L121.}

\ref{Wurm, G. 1997. PhD Thesis. University of Jena.}

\ref{Wurm, G., and Blum, J. 1998. Experiments on preplanetary dust aggregation. {\refit Icarus\/} 132:125-136.}

\ref{Yoneda S., Grossman L. 1995. Condensation of CaO-MgO-Al$_2$O$_3$-SiO$_2$ liquids from cosmic gas. {\refit Geochim. Cosmochim. Acta\/} 59:3413-3444.}

\bye